\documentstyle[12pt]{article}
\newcommand{\be}{\begin{equation}}
\newcommand{\ee}{\end{equation}}
\newcommand{\s}{\section}
\newcommand{\ci}{\cite}

\begin{document}

\begin{titlepage}

\begin{flushright}
FTUV/92-27
\end{flushright}

\begin{center}

\vspace{2.0cm}

{\Large{\bf Pionic effects in deep inelastic scattering off nuclei  $^*$}}\\

\vspace{1.7cm}

{\large P. Gonz\'alez $^{**}$ and V.Vento $^{\ddag}$}\\

\vspace{0.2cm}
{Departament de F\'{\i}sica Te\`{o}rica and I.F.I.C.}\\
{Centre Mixt Universitat de Val\`{e}ncia -- C.S.I.C.}\\
{E-46100 Burjassot (Val\`{e}ncia), Spain.}

\vspace{2.0cm}
{\bf Abstract}
\vspace{0.2cm}
\begin{quotation}
{\small The structure functions calculated in the Chiral bag
model reproduce quite well, after appropriate perturbative
evolution to large energy scales, the experimental data. We use
these results to interpret the structure of the $EMC$ data
as a quenching of the pion decay constant due to the in medium
behavior of the nucleon. This explanation supports recent
proposals of this phenomenon whose origin is the scale invariance
of the $QCD$ lagrangian.}
\end{quotation}
\end{center}
\vspace{3.5cm}
$^*$Supported in part by CICYT grant \# AEN90-0040 and DGICYT
grant \# PB88-0064. \\
$^{**}$ Gonzalep@evalvx.bitnet\\
$^{\ddag}$ Vento@vm.ci.uv.es

\end{titlepage}

\baselineskip  0.2in

\s{Introduction}

The European Muon Collaborations ($EMC, NMC,SMC$) have been
shaking the foundations of our knowledge of the nucleon structure
for the past ten years \ci{emc,nmc}. In this paper we shall reanalyze
the data of the original $EMC$ effect and find room for an
heterodox explanation associated with some recent proposal about
the scaling properties of coupling constants in a hadronic medium
\ci{br}.

In order to explain the $EMC$ data, two broad classes of models
have been developed \ci{ro}. One of the approaches comprises
conventional nuclear physics models and dynamics. The contributions
of all scattering components of the nucleus are taken into account,
so in addition to the traditional protons
and neutrons, constituent mesons and nucleon resonances have been
considered \ci{np}. Furthermore the properties of the nuclear
medium, i.e. binding energy and density, can be incorporated
\ci{vu}.

A second approach is based on the so called $QCD$ rescaling models. Here
it is assumed that the internal $QCD$ dynamics is modified in the
nuclear environment leading to changes in the dynamics which
can be described by associating a different renormalization group
scale to each nuclei \ci{vm}. This mechanism interpreted in terms
of a bag model has led to the much discussed phenomenon of
nucleon {\em swelling}.

We pursue in this letter a different philosophy, namely we would
like to establish a link, using experimental high energy data,
between effective low energy models and asymptotic properties of
the theory. This approach, originally proposed by Jaffe and Ross
\ci{jr}, has been followed by other groups as well \ci{thb}.
In this spirit we have performed calculations of the structure
functions in the Chiral bag model for single nucleon
properties and extended them to incorporate medium effects.

\s{The structure function of the nucleon}

The first step in our approach is the calculation of the structure
functions in a conventional low energy model. We have chosen for
this purpose the chiral bag model, a model described in terms of
quarks and gluons fields in the interior of a cavity, the bag,
and in terms of pion fields in its exterior, while their coupling is
defined to implement the symmetries of the strong interactions, color
gauge symmetry and spontaneously broken chiral symmetry \ci{vv}.
Following conventional techniques we calculate Compton scattering
in the model by using the chiral expansion to first order in
$\frac{1}{4 \pi f_{\pi}^2R^2}$, the effective expansion
parameter. We obtain three types of terms :
i) the photon-quark (antiquark) scattering inside the cavity, the
only contribution appearing in the original
MIT bag model calculation \ci{jam}; ii) the photon-quark (antiquark)
scattering with pionic final state interactions, arising
automatically in the chiral expansion to this order; iii) the
photon-(virtual)pion interaction, corresponding to the meson
exchange current contributions in the model. Only in this latter
type of diagrams the photon scrutinizes, due to its short
wave length, the structure of the boson.
In order to connect with the asymptotic regime (only quarks,
antiquarks and gluons) we incorporate, only in the meson exchange
contribution, the structure of the pion \ci{sv}. In this way the
model differs from the more naive convolution models \ci{np} by
distinguishing among the pionic contributions those with different
time scales \ci{jac,thp}. In particular those diagrams leading to
final state quark-gluon interactions where a quark from the exchanged
pion interacts with a quark from the nucleon by means of gluon exchange,
which could change our internal exchanged pion time-scale,
do not exist in our two phase approach.
Between two different
hadrons only pions can be exchanged and the relevant contributions are
incorporated in the pionic description of the final state
interaction. Certainly the model is very naive
describing the final state interactions, but we would like to
stress that, within an effective chiral
theory approach, ours is the only possible calculation to lowest order
in the chiral expansion.

Following Jaffe and Ross \ci{jr}, radiative corrections are
considered by adscribing a low energy scale, $Q_0$, to the model calculation
and evolving to high $Q$ using the renormalization group
equations. This procedure is certainly justified for the twist
two operators, which are the only ones we are interested in here.

\begin{figure}[t]
\vspace{7cm}
\caption{The isoscalar structure function of the nucleon.
The solid line represents the structure function
calculated in the Chiral bag model for $ R = 1 fm, f_\pi = 95
MeV$. The dashed line represents its evolved from
$\frac{Q_0}{\Lambda_{QCD}} = 1.5 $ to $\frac{Q}{\Lambda_{QCD}} =
25$. Finally the vertical lines represent an eye guide to data points
of the different experiments and their errors [18].}
\label{sf}
\end{figure}

There has been a lot of discussion about the so called support
problem \ci{ja,ths}. It is certainly true that bag calculations
have a difficult center of mass problem which leads to structure
functions which are not limited to the interval $0 \le x \le 1$,
where $x$ is the Bjorken variable. However in most calculations
over 95 \% of the momentum is inside that interval, as is our
case. The support problem cannot be solved exactly, except in the
two dimensional case. Therefore many approximations have been
developed based on the calculations of quark distribution
functions followed by a Peierls-Yoccoz projection. Due to the
relativistic character of bag model calculation, this procedure
is not only approximate, but very hard to implement. Moreover
technical difficulties force one to severely restrict the
number of modes in the calculation of the cavity propagators.
In our case, the Chiral bag model, where one has to take also into
account the pionic degrees of freedom, its
development is tedious, technically complex and the errors due to
truncation not controlable.
In ref.(\ci{thp}) the projection procedure was followed, but
the calculation of terms to which the pion propagator could
contribute was eluded.

Since our calculation was planned to unveil new dynamical
possibilities, namely the implication of scale invariance into
effective low energy actions, we have chosen simplicity, where matters were
difficult to control, i.e., we have used unprojected states.
On the contrary, whenever the scheme was well defined, we have
not avoided calculational effort and have aimed to the highest
precission possible. For example,  we have carried
out a quite complete (up to 50 modes have been considered) cavity
perturbation theory calculation in the chiral bag including all
the pionic diagrams.

We have forced our structure function to have the proper
support, by restricting $x$ to vary in the correct interval
($[0,1]$). We have analyzed the
stability of the solution by enlarging or reducing
the support about these endpoints. The stability of our results
after truncation give us some confidence in the qualitative
behavior of the evolved structure function obtained.

We have used an ingenious method of evolution \ci{g-a} to leading
order which requires relatively few moments: from the tenth moment on
results do not change more than 15 \% for $x > 0.05$. The
truncation of the support affects mostly the high moments.
Recall that it is precisely the high moments which might create
disturbances in the calculation due to the ill-behavior of the
structure function beyond $x = 1$.
This reconstruction method is therefore greatly responsible for
the stability of our solution.
However one has to be extremely careful performing the
calculation since rounding errors can destroy the reconstruction
procedure altogether. The method requires the operation with very large
numbers to obtain relatively small results.
In our case, a careful optimization of the
calculational sequence has allowed us to compute
up to a maximum of 35 moments. Beyond this number we run into precission
problems. The the method was generalized to a different support
in order to check convergence.

In Fig.(\ref{sf}) we show the Chiral bag model structure function
and the one that results from our evolution process. The
parameters of the model have been fixed in part to reproduce the
experimental low energy data, without much effort at precission
(bag radius $R = 1 \; fm$, $f_{\pi} = 95\; MeV$, which lead
to a mass for the nucleon of $M \approx 1000 \;MeV$). To leading order the
remaining
parameter is $\frac{Q_0}{\Lambda_{QCD}}$, which we have fitted
to get the correct behavior of the structure function and the
maximum possible value for the gluon content. Our value for it of
$1.5$ is consistent with that of other calculations \ci{thb}.
{}From this ratio and an {\em experimental} $\Lambda_{QCD} = 0.2\; GeV$
the  low energy scale becomes $Q_0 = 0.3\; GeV$.  We have
evolved up to $Q = 5 \;GeV$, although one reaches a slowly
varying plateau in the evolution from $Q = 1.5\; GeV$ on.
It is clear from the figure that we overestimate the low $x$
behaviour of the structure function. On the other hand the gluon
content of the evolved solution is about 47 \% instead of the
experimental one of 55 \%.

These two features seem to indicate that we should
add to our description hard gluons at the quark model level
through the one gluon exchange mechanism ($OGE$) \ci{thb,mth}.
In this case one would have gluons at the low evolution scale,
which would also imply an increase in the
$\frac{Q_0}{\Lambda}$ ratio and as a consequence a shift of
the momentum from the low $x$  to intermediate $x$ quarks,
approaching a overall better fit to the data.

We have calculated, with the additional assumption of a symmetric
sea, the Drell-Yan process leading to a fit of similar quality.
The gluon distribution comes out comparable to other more
phenomenological approaches \ci{ro}, but the full merit here
corresponds to the $QCD$ evolution  and not to the model.

\s{The structure function in a nuclear medium}

By using effective chiral Lagrangians with a suitable
incorporation of the scaling property of $QCD$, Brown and Rho
\ci{br} have established an in-medium scaling law one of whose
most appealing expressions reads
\be
\frac{M^*}{M} =
\sqrt{\frac{g^*_A}{g_A}} \;\frac{f^*_{\pi}}{f_{\pi}}
\label{brm*}
\ee
where $M^*$ , $g^*_A$ and $f_{\pi}^*$ are the
effective in-medium nucleon mass, axial coupling constant and
pion decay constant respectively. The ratio of pion decay constants,
our scaling
parameter, will be labeled hereafter $\sigma$. We follow their
prescription and assume that the in-medium effects can be
incorporated by taking the effective Lagrangian with scaled
couplings to calculate physical observables according to a chiral
perturbation scheme. In our case this reduces to a chiral bag model
picture in which the quarks, gluons and pions carry effective charges.

In our scheme, to lowest order in the
chiral expansion, $g_A$ does not change in the medium, i.e.,
\be
g_A = \frac{5}{9}\frac{\varepsilon}{\varepsilon - 1}
\ee
where $\varepsilon = 2.04...$ is the lowest mode of the cavity,
which does not change with density.

This is not of major concern since experimentally
$\sqrt{\frac{g^*_A}{g_A}} > 0.9$
and quite close to one for intermediate nuclei, which are the
ones used in EMC data. Moreover in our calculation the structure
function is strictly independent
on the mass as long as we adhere to the chiral
perturbative scheme to lowest order (It turns out to be  only a function of
$\varepsilon$ and $f_{\pi}R$).
Our scheme will produce a dependence of the structure function on
$MR$ and of $g_A$ on $MR$ and $f_{\pi}R$ in
higher orders in the chiral expansion \ci{rho}. We shall not consider
these effects in the present work and hence the
only effective medium dependent parameter to be considered is $f_{\pi}$.

In Fig.(\ref{emc})
we plot the ratio $\frac{F_2(f^*_{\pi})}{F_2(f_{\pi})}$ and
compare it with the $EMC$ data. Certainly we are aware of Fermi
motion corrections \ci{ro,vu}, but to clarify our discussion we
have not included them in our calculation.

\begin{figure}[t]
\vspace{7cm}
\caption {Ratio of structure functions for different values of the
scaling parameter. The continuous curve corresponds to $\sigma =
.95$ and the dashed curve to $\sigma = 0.9$. The vertical lines
give a schematic idea of the spread of data points of the $EMC$
effect for the different experiments and their errors [18].}
\label{emc}
\end{figure}
Our aim has not been to obtain a one parameter ($\sigma$) fit to the
experimental EMC data, although from the results shown in
Fig.(\ref{emc}) one could attempt to do so incorporating the
missing ingredients (support corrections, loop contribution, fermi
motion contributions, etc...). We want to signal however, that the proposed
mechanism is quite efficient in shifting momentum from the
intermediate $x$ region to the low $x$ region, and that the data
seem to support such a quenching of the effective parameters.

{}From our calculation we have also obtained the gluon
distribution. Our mechanism does produce a change of it in
going from free nucleons to in-medium nucleons, which is
similar in shape to the structure function ratio, but smaller
in magnitude. As in the free case, since gluons are produced
solely via bremsstrahlung, the result is very much dependent
on the evolution ratio.

Before finishing this section one caveat is necessary. The
reader might find surprising that our data does not extend into
the shadowing region. As is well known, shadowing has to do with
the hadronic structure of the photon, and we have never aimed at
discussing this physics. Therefore the very low $x$ data could
only confuse the issue.

\s{Conclusion}

We have performed a calculation of the structure function of the
nucleon using as boundary conditions of the renormalization group
equations the structure functions calculated in the Chiral bag
model. A qualitative fit to the experimental data is obtained
with reasonable values of the parameters. In particular small
value of the low energy scale parameter $Q_0$ may have to do with
the omission of perturbative gluonic contributions to the
structure function, relevant at this scale where the
$QCD$ coupling constant is sizeable. The smallness of $Q_0$
leads to an overestimate of the structure
function for low $x$, since the evolution equations in this case
shift too much quark momentum from the high to the low $x$ region.

Using the scaling argument of Brown and Rho, we have been able to
obtain the in-medium structure functions. We see that the
experimental results are compatible with such a mechanism.
The quenching of $f_{\pi}$ is relatively small, as corresponds
to low densities. Moreover once Fermi motion corrections are
incorporated the data can be well fitted. Our aim, however is not to
obtain a one parameter fit to the data, but to reveal that the
scaling hypothesis could be confirmed in this scenario.

Peculiar features of the calculation are the rise for low
$x$ and high $x$ of the ratio as can be seen in Fig.(\ref{emc}).
The behavior at low $x$ is in our opinion a general
characteristic of pionic contributions and will appear in any
model calculation including them. It arises from the meson
exchange diagram and is a direct consequence of the structure
of the pion \ci{sv}. Certainly this rise at low x carries,
due to momentum conservation
a depletion of the intermediate $x$ region. On the contrary the
rise at high $x$ is a peculiarity of the model. The confinement
condition in the Chiral bag model leads to boundary conditions
originating a quark pion interaction which becomes stronger as we
decrease  $f_{\pi}$. This interaction determines the strenght and
$x$-dependence of the final state interaction terms
(those of type ii)) producing for the scaled parameter
an inverse shift pushing quarks from
intermediate $x$ to high $x$ \ci{sv}.

We must end by stating that in our opinion despite
the technical difficulties
that remain to be solved the approach used in this letter to
signal a possible scaling scenario deserves to be further
developed. The behavior of the structure function in
Fig.(\ref{sf}), together with the results of other calculations
\ci{thb,thp},  makes the whole process appealing. Moreover we see
that one can implement a low energy microscopic description by
using high energy data, as long as one accepts the evolution
assumption \ci{jr}.

\s*{Acknowledgement}
We acknowledge useful conversations with Marco Traini, Jos\'e
Pe\~narrocha and V. Sanjos\'e. P.G. is thankful to Celia for
assistance in the optimization of the computer codes.


\begin{thebibliography}{99}
\bibitem{emc}
J.J. Aubert, Phys. Lett. {\bf 123B} (1983) 275.
\bibitem{nmc}
J. Ashman et al., Phys. Lett. {\bf B206} (1988) 364; P. Amaudruz
et al., Phys. Rev. Lett. {\bf 66} (1991) 2712.
\bibitem{br}
G.E. Brown and M. Rho, Phys. Rev. Lett. {\bf 66} (1991) 2720.
\bibitem{ro}
R.G. Roberts in The structure of the proton (Cambridge University
Press 1990); R.D. Field in Applications of perturbative $QCD$
(Addison Wesley Pub. Co. 1989;
V. Barone and E. Predazzi, Ann. Phys. (Fr.) {\bf 12} (1987) 525.
\bibitem{np}
C.H. Llewelyn Smith, Phys. Lett. {\bf 128B} (1983) 107; M.
Ericson and A.W. Thomas, Phys. Lett. {\bf 128B} (1983) 112; J.
Szwed, Phys. Lett {\bf 128B} (1983) 245.
\bibitem{vu}
C.A. Garcia Canal, E.M. Santangelo and H. Vucetich, Phys. Rev.
Lett. {\bf 53} (1984) 1430; Phys. Rev. {\bf D35} (1986) 382.
\bibitem{vm}
R.L. Jaffe, F.E. Close, R.G. Roberts and G.G. Ross, Phys. Lett.
{\bf 134B} (1984) 449; Phys. Rev. {\bf D31} (1985) 1004.
\bibitem{jr}
R.L. Jaffe and G.G. Ross, Phys. Lett {\bf 93B} (1980) 313.
\bibitem{thb}
A.W. Schreiber, A.I. Signal and A.W. Thomas, Phys. Rev. {\bf D44}
(1991) 2653; L. Conci and M. Traini, Few-Body Systems {\bf 8}
(1990) 123; M. Traini, L. Conci and U. Moschella, Nucl. Phys. {\bf A544}
(1992) 731; M. Traini, L. Conci and M. Melchiori, University of
Trento UTF 237-1991; V. Barone and A. Drago, University of Ferrara 1992.
\bibitem{vv}
V. Vento, M. Rho, E. Nyman, J.H. Jun, and G.E. Brown, Nucl. Phys.
{\bf A470} (1980) 413.
\bibitem{jam}
R.L. Jaffe, Phys. Rev. {\bf D11} (1975) 1953.
\bibitem{sv}
V. Sanjos\'e, V. Vento and S. Noguera, Nucl. Phys. {\bf A470}
(1987) 510; V. Sanjos\'e and V. Vento, Nucl. Phys. {\bf A501}
(1989) 672.
\bibitem{jac}
R.L.Jaffe, Comments in Nucl. Part. Phys. {\bf 13} (1984) 39.
\bibitem{thp}
A.W. Schreiber, P.J. Mulders, A.I. Signal and A.W. Thomas, Phys.
Rev. {\bf D45} (1992) 3069.
\bibitem{ja}
R.L. Jaffe, Ann. Phys. (N.Y.) {\bf 132} (1981) 32.
\bibitem{ths}
A.I. Signal and A.W. Thomas, Phys. Rev. {\bf D40} (1989) 2832.
\bibitem{g-a}
F.J. Yndur\'ain, Phys. Lett. {\bf B74} (1978) 68;
A. Gonz\'alez-Arroyo, C. L\'opez and F.J. Yndur\'ain, Nucl. Phys.
{\bf B153} (1979) 161; ibid {\bf B159} (1979) 512; ibid {\bf
B166} (1980) 429.
\bibitem{rpp}
J.J. Hern\'andez et al., Phys. Lett. {\bf B239} (1990) 1;
R.G. Roberts and M.R. Whalley, J. Phys. G {\bf 17} (1991) D1.
\bibitem{mth}
F. Myhrer and A.W. Thomas, Phys. Rev. {\bf D38} (1988) 1633.
\bibitem{rho}
M. Rho, Phys. Rev. Lett. {\bf 54} (1985) 767; G. E. Brown and M.
Rho, Phys. Lett. {\bf B 222} (1989) 324.
\end{thebibliography}
\end{document}